%% file: main.tex
\documentclass{article}
\usepackage[preprint]{spconf}
\usepackage{amsmath,amssymb,graphicx,hyperref}
\usepackage{subcaption}
\usepackage{cite}
\usepackage{acronym}
\usepackage{amsthm}
\usepackage[table]{xcolor}
\usepackage{multirow}
\usepackage{multicol}
\usepackage{booktabs}
\usepackage{bm}
\usepackage[inline]{enumitem}
\usepackage{etoolbox}
\usepackage{xurl}
\usepackage[protrusion=true, expansion=true, tracking=small]{microtype}

\newtoggle{preprint}

\toggletrue{preprint}

\newcommand{\mathvar}[2]{\newcommand{#1}{\ifmmode{#2}\else$#2$\fi}}

\acrodef{CM}{Consistency Model}
\acrodef{CT}{Consistency Training}
\acrodef{CAE}{Consistency Autoencoder}
\acrodef{DDPM}{Denoising Diffusion Probabilistic Model}
\acrodef{SDE}{Stochastic Differential Equation}
\acrodef{ODE}{Ordinary Differential Equation}
\acrodef{GAN}{Generative Adversarial Network}
\acrodef{VAE}{Variational Autoencoder}
\acrodef{EMA}{Exponential Moving Average}
\acrodef{iCT}{improved Consistency Training}
\acrodef{SSL}{Self-Supervised Learning}
\acrodef{VQ}{Vector Quantization}
\acrodef{AE}{Autoencoder}
\acrodef{DiffAE}{Diffusion Autoencoder}

\acrodef{DSP}{Digital Signal Processing}
\acrodef{STFT}{Short-Time Fourier Transform}
\acrodef{SNR}{Signal-to-Noise Ratio}
\acrodef{ISTFT}{Inverse Short-Time Fourier Transform}

\acrodef{SNR}{Signal to Noise Ratio}
\acrodef{SDR}{Signal to Distortion Ratio}
\acrodef{SI-SDR}{Scale-invariant Signal to Distortion Ratio}
\acrodef{MSS}{Multi-Scale Spectral distance}
\acrodef{FAD}{Fréchet Audio Distance}
\acrodef{KAD}{Kernel Audio Distance}
\acrodef{ERank}{Effective Rank}

\newtheorem{property}{Property}

\newcommand{\sectionsep}{0.1cm}

\mathvar{\denoiser}{f_\theta}              
\mathvar{\encoder}{g_\theta}
\mathvar{\network}{F_\theta}              
\newcommand{\Enc}{\operatorname{Enc}_\theta}         
\newcommand{\Dec}{\operatorname{Dec}_\theta}         
\mathvar{\distance}{d}                    

\mathvar{\xclean}{x}                      
\mathvar{\yclean}{y}                      
\mathvar{\xsigma}{x_\sigma}               
\mathvar{\latx}{\text{lat}_x}             
\mathvar{\Z}{\mathbf{z}}           
\mathvar{\X}{\mathbf{x}}
\mathvar{\Y}{\mathbf{y}}
\mathvar{\U}{\mathbf{u}}
\mathvar{\V}{\mathbf{v}}
\mathvar{\xhat}{\hat{x}}                  

\mathvar{\noise}{\sigma}                  
\mathvar{\noisei}{\sigma_i}               
\mathvar{\noiseinext}{\sigma_{i+1}}        

\mathvar{\params}{\theta}                 
\mathvar{\paramsema}{\theta^-}            
\mathvar{\cskip}{c_{\text{skip}}}         
\mathvar{\cout}{c_{\text{out}}}           
\mathvar{\lambdaweight}{\lambda}          

\mathvar{\Lct}{\mathcal{L}_{\text{CT}}}   
\mathvar{\E}{\mathbb{E}}                  
\mathvar{\R}{\mathbb{R}}                  
\mathvar{\lhuber}{\mathcal{L}_{\text{huber}}} 
\mathvar{\C}{\mathbb{C}}

\mathvar{\constc}{c}                      
\mathvar{\ctilde}{\tilde{c}}              
\mathvar{\phasec}{\angle(c)}              
\mathvar{\amplitudetransform}{\operatorname{Amp}} 
\mathvar{\STFT}{\text{STFT}}  
\mathvar{\T}{\operatorname{T}} 

\newcommand{\EPS}{\bm{\epsilon}}               

\iftoggle{preprint}{
  \mathvar{\reda}{{\color{magenta}a}}
  \mathvar{\blueZx}{{\color{blue}\Z_x}}
  \mathvar{\blueZ}{{\color{blue}\Z}}
  \mathvar{\blueZy}{{\color{blue}\Z_y}}
  \mathvar{\Zxy}{\mathbf{Z}_{u+v}}
  \mathvar{\blueZxplusy}{{\color{blue}\Z_{u}+\Z_v}}
  \mathvar{\blueZxy}{{\color{blue}\Z_{u+v}}}
  \newcommand{\blue}[1]{{\color{blue} #1}}

}{
  \mathvar{\reda}{a}
  \mathvar{\blueZx}{\Z_x}
  \mathvar{\blueZ}{\Z}
  \mathvar{\blueZy}{\Z_y}
  \mathvar{\Zxy}{\mathbf{Z}_{u+v}}
  \mathvar{\blueZxplusy}{\Z_{u}+\Z_v}
  \mathvar{\blueZxy}{\Z_{u+v}}
  \newcommand{\blue}[1]{#1}
}

\title{Learning Linearity in Audio Consistency Autoencoders via Implicit Regularization}
\name{Bernardo Torres$^{\star}$\thanks{$\star$Work done during an internship at Deezer Research} \qquad Manuel Moussallam$^{\dagger}$ \qquad Gabriel Meseguer-Brocal $^{\dagger}$}

\address{\small
         $^{\star}$LTCI, Telecom Paris, Institut Polytechnique de Paris \\
         \small $^{\dagger}$Deezer Research}
\copyrightnotice{\copyright 2026 IEEE. Personal use of this material is permitted. Permission from IEEE must be obtained for all other uses, in any current or future media, including reprinting/republishing this material for advertising or promotional purposes, creating new collective works, for resale or redistribution to servers or lists, or reuse of any copyrighted component of this work in other works.}

\begin{document}
\ninept

\maketitle

\begin{abstract}
  Audio autoencoders learn useful, compressed audio representations, but their non-linear latent spaces  prevent intuitive algebraic manipulation such as mixing or scaling. We introduce a simple training methodology to induce linearity in a high-compression Consistency Autoencoder (CAE) by using data augmentation, thereby inducing homogeneity (equivariance to scalar gain) and additivity (the decoder preserves addition) without altering the model's architecture or loss function. When trained with our method, the CAE exhibits linear behavior in both the encoder and decoder while  preserving reconstruction fidelity. We test the practical utility of our learned space on music source composition and separation via simple latent arithmetic. This work presents a straightforward technique for constructing structured latent spaces, enabling more intuitive and efficient audio processing.

\end{abstract}
\section{Introduction}\label{sec:intro}

Modern \acp{AE} can achieve excellent reconstruction quality at high compression rates at the expense of complex, entangled latent spaces. For applications where input space manipulation is desirable from within the compressed space, recent research has proposed either task-specific adaptors for pre-trained models or redesigning the autoencoder to preserve key structural properties, such as equivariance to spatial transformations \cite{kouzelisEQVAEEquivarianceRegularized2025,skorokhodovImprovingDiffusabilityAutoencoders2025, zhouAliasfreeLatentDiffusion2025}. This work follows the latter approach.

\begin{figure}
    \centering
    \includegraphics[width=0.75\columnwidth, trim={10 12 10 10}]{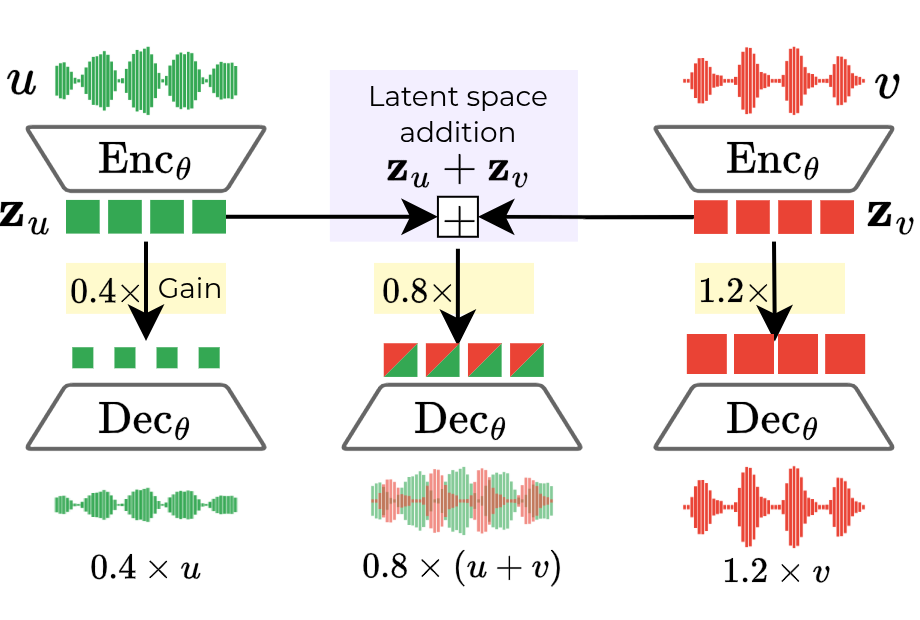}
    \caption{In a linear decoder, applying a gain to the latent vector scales the output by the same gain (homogeneity), and summing  latents corresponds to a sum in the audio domain (additivity).}
    \label{fig:overview}
  \end{figure}

For certain applications in audio, linearity (Figure \ref{fig:overview}) is a desirable property. A linear map fulfills two properties: \begin{enumerate*}[label=(\Roman*)]
\item Homogeneity: scaling the input by a value scales the output by the same value; and 
\item Additivity: the map preserves addition.
\end{enumerate*}
As processing large audio datasets in the latent space becomes more prevalent, direct mixing and volume adjustment in this space can improve efficiency by reducing redundant encoding and decoding. Additionally, as downstream tasks such as audio generation and source separation can benefit from composition via latent arithmetic, better interpretability may be achieved when working on a linear space.

This work presents a training methodology for constructing an approximately linear compressed audio representation that enables intuitive manipulation, where simple algebraic operations in the latent space correspond directly to mixing and scaling in the audio domain. The method employs implicit regularization through data augmentation, without modifying the model architecture or objective. The approach is demonstrated with the Music2Latent architecture \cite{pasiniMusic2LatentConsistencyAutoencoders2024}, a \ac{CAE} that achieves high-quality, single-step reconstruction and a $64\times$ compression rate for $44.1$ kHz audio.

Our main contributions are:
\begin{enumerate*}[label=(\Roman*)]
    \item An unsupervised, data-augmentation-based training procedure that induces approximate linearity in a high compression \ac{AE}, with no extra loss terms;
    \item Validation on a state-of-the-art \ac{CAE} for music and speech, showing linearity in both encoder and decoder with no loss of reconstruction quality;
    \item Practical utility shown on oracle source separation via simple latent arithmetic.
\end{enumerate*} 
Code and model weights are available online\footnote{\scriptsize{\url{www.github.com/bernardo-torres/linear-autoencoders}.}}. 

\begin{figure*}[!ht]
  \centering
  
  \newdimen\totalheight
  \setlength{\totalheight}{5cm}

  \begin{minipage}[c]{0.6\textwidth}
    \centering
    \newsavebox{\subfigboxone}
    \sbox{\subfigboxone}{\includegraphics[height=\totalheight, trim={0 6 0 25}]{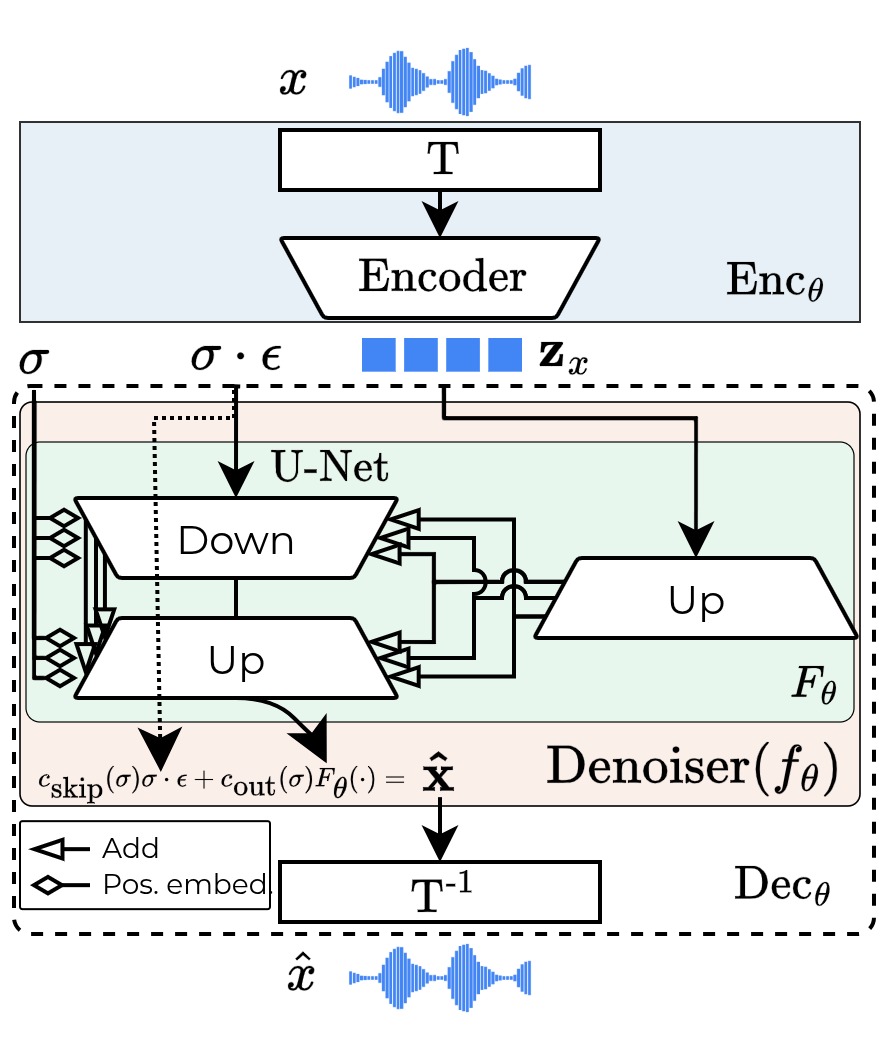}}
    \begin{subfigure}[c]{\wd\subfigboxone}
        \usebox{\subfigboxone}
        \caption{Music2Latent}
        \label{fig:part1}
    \end{subfigure}
    \hfill
    \newsavebox{\subfigboxtwo}
    \sbox{\subfigboxtwo}{\includegraphics[height=\totalheight, trim={-3 6 -3 25}]{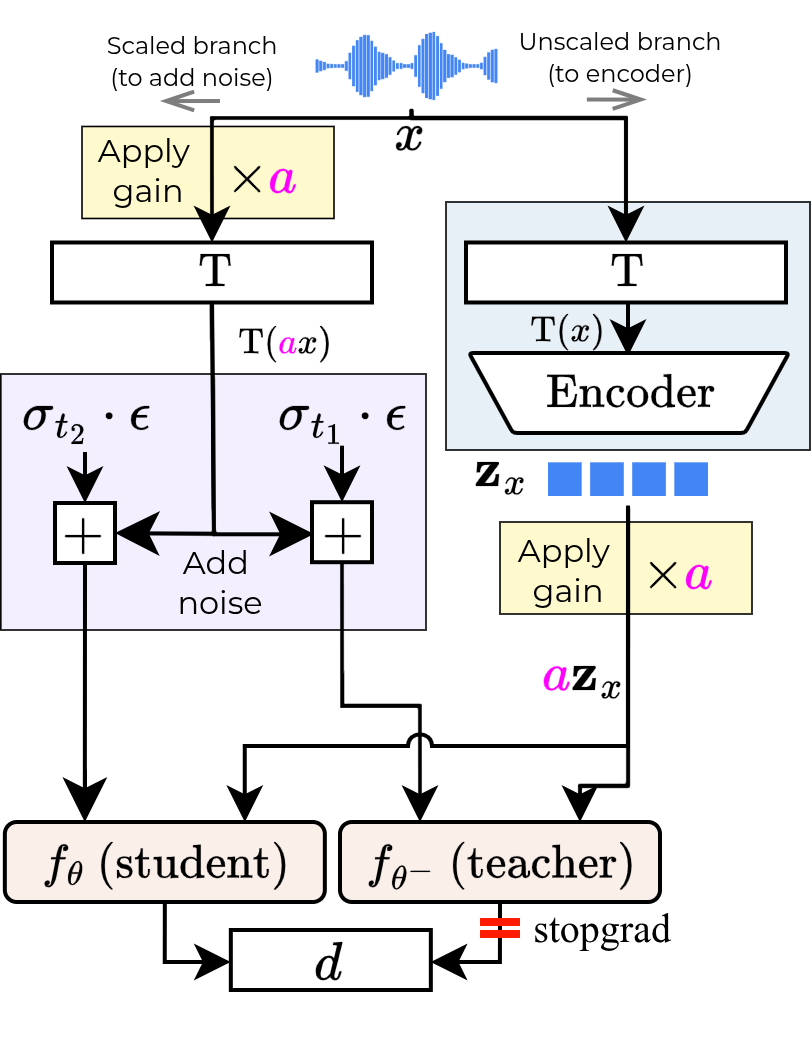}}
    \begin{subfigure}[c]{\wd\subfigboxtwo}
        \usebox{\subfigboxtwo}
        \caption{\ac{CAE} training with implicit homogeneity}
        \label{fig:part2}
    \end{subfigure}
  \end{minipage}%
  \hfill
  \begin{minipage}[c]{0.4\textwidth}
    \centering
    \newsavebox{\subfigboxthree}
    \sbox{\subfigboxthree}{\includegraphics[height=0.5\totalheight, trim={0 0 0 30}]{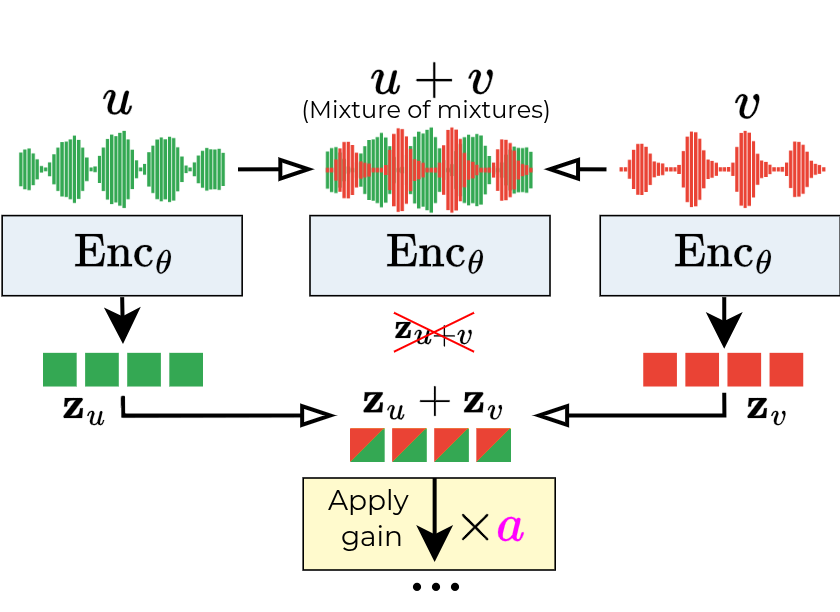}}
    \begin{subfigure}[c]{\linewidth}
        \centering
        \usebox{\subfigboxthree}
        \caption{\ac{CAE} training with implicit additivity}
        \label{fig:part3}
    \end{subfigure}
    \vfill
    \newsavebox{\subfigboxfour}
    \sbox{\subfigboxfour}{\includegraphics[height=0.4\totalheight, trim={0 5 0 0}]{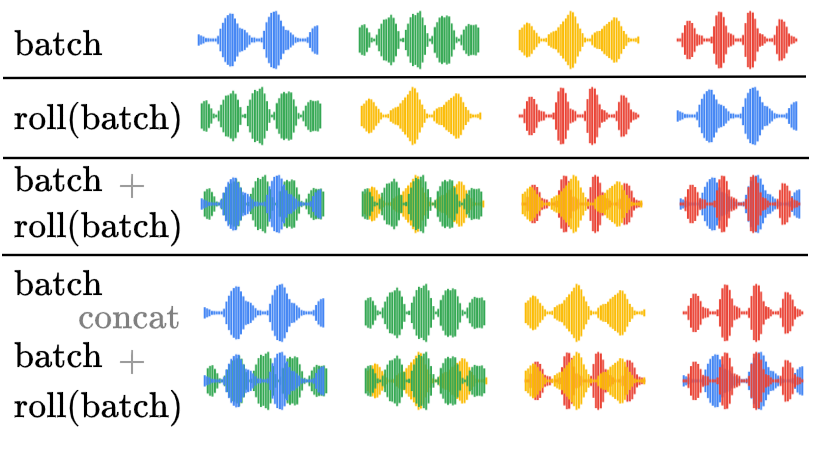}}
    \begin{subfigure}[c]{\linewidth}
        \centering
        \usebox{\subfigboxfour}
        \caption{Batch creation}
        \label{fig:part4}
    \end{subfigure}
  \end{minipage}
  
  \caption{(\subref{fig:part1}): Music2Latent \ac{CAE} architecture. The decoder is a denoising U-Net and the latent is introduced to it at every resolution level after learned upsampling. (\subref{fig:part2}): Proposed  \ac{CAE}  training trick to implicitly enforce homogeneity in the decoder. (\subref{fig:part3}): Proposed trick to enforce additivity, applied when the input is an artificial mixture. (\subref{fig:part4}): Batch creation procedure with artificial mixtures of mixtures. }
  \label{fig:diagram}
\end{figure*}


\section{Background}

\subsection{Audio processing in the latent space}

There is a growing interest in manipulating audio by training task-specific modules which operate directly in the latent space of a pretrained \ac{AE}, with applications in source separation \cite{mancusiUnsupervisedSourceSeparation2021,postolacheLatentAutoregressiveSource2023, bindiUnsupervisedComposableRepresentations2024}, speech enhancement \cite{omranDisentanglingSpeechSurroundings2023, liSpeechEnhancementUsing2025}, upsampling and upmixing \cite{braliosLearningUpsampleUpmix2025}, filtering \cite{braliosRebottleneckLatentRestructuring2025}, and  generative modelling \cite{schneiderMousaiEfficientTexttomusic2024,evansStableAudioOpen2025, nistalDiffariffMusicalAccompaniment2024}.
Some works retrain \acp{AE} for specific objectives, such as source disentanglement \cite{bieLearningSourceDisentanglement2025} or separation via latent masking \cite{tzinisTwostepSoundSource2020}. In contrast, our work focuses on training an \ac{AE} to have desirable, task-agnostic structural properties. This enables direct and efficient audio manipulation and can serve as a strong foundation for these downstream applications.

\subsection{Diffusion and Consistency models}

Denoising Diffusion Probabilistic Models \cite{hoDenoisingDiffusionProbabilistic2020a} and score-based models \cite{songScorebasedGenerativeModeling2021}  have achieved great success in generative modeling, where the goal is to estimate the underlying data distribution from samples.
 These models define a forward process that gradually adds noise to data and learn a reverse process to denoise it using a neural network. Sampling typically requires an iterative procedure with many steps to generate a clean sample. \acp{CM} \cite{songConsistencyModels2023} accelerate this process by mapping any point along the trajectory defined by the probability flow ordinary differential equation \cite{songScorebasedGenerativeModeling2021} directly to the origin (the clean data point), enabling single-step generation. When trained from scratch, the process is called \acf{CT}, in which a "student" denoiser network (\denoiser) is trained to match the output of a "teacher" ($f_{\theta^{-}}$), which itself denoises a less corrupted version of the same data point. In \ac{iCT} \cite{songImprovedTechniquesTraining2023}, the teacher is updated with the same parameters $\theta$ as the student but detached from the computational graph  ($\theta^- \gets \text{stopgrad}(\theta)$).

\vspace{-0.2cm}

\subsection{High-Fidelity Audio Autoencoders}

In audio, popular \acp{AE} include Neural Audio Codecs \cite{defossezHighFidelityNeural2023}, which compress audio into a set of discrete tokens, and \acp{VAE} \cite{kingmaAutoencodingVariationalBayes2014}, which have seen significant success in generative modeling. However, both frameworks require complex, multi-stage training with adversarial objectives \cite{esserTamingTransformersHighresolution2021} to achieve high-quality reconstructions. \acp{DiffAE} \cite{preechakulDiffusionAutoencodersMeaningful2022,birodkarSampleWhatYou2024, chenDiffusionAutoencodersAre2025,zhaoEpsilonvaeDenoisingVisual2024, schneiderMousaiEfficientTexttomusic2024} enable high-fidelity reconstruction by replacing the deterministic decoder with a conditional diffusion model. Notable examples in audio include \cite{schneiderMousaiEfficientTexttomusic2024} and Music2Latent \cite{pasiniMusic2LatentConsistencyAutoencoders2024,pasiniMusic2Latent2AudioCompression2025}. A specific instance of \acp{DiffAE} is the \ac{CAE}\cite{pasiniMusic2LatentConsistencyAutoencoders2024,pasiniMusic2Latent2AudioCompression2025}, where the decoder is a \ac{CM}, enabling decoding in one step. \acp{DiffAE} can be trained with a single diffusion/\ac{CT} objective \cite{preechakulDiffusionAutoencodersMeaningful2022, chenDiffusionAutoencodersAre2025, pasiniMusic2LatentConsistencyAutoencoders2024}. High-quality decoding has been linked to the capacity of sampling details at inference-time, since the decoder acts as a denoiser instead of an upsampler, which also reduces the amount of unnecessary information to be encoded in the latent \cite{birodkarSampleWhatYou2024, zhaoEpsilonvaeDenoisingVisual2024}.
\section{Method}

Let $\mathcal{M} \subset [-1, 1]^T$ be the space of audio signals of length $T$ of interest and $\mathcal{Z} \subset \R^{N \times F}$ be a lower-dimensional latent space induced by encoder $\Enc: \mathcal{M} \to \mathcal{Z}$ and decoder $\Dec: \mathcal{Z} \to \mathcal{M}$, with dimensions $N, F$ defined by the compression factor.
 We train $(\Enc, \Dec$) with \ac{CAE} training under the constraint that $\Dec$ is approximately linear (Figure \ref{fig:overview}), i.e., it satisfies the following properties:
 
\begin{property}[Homogeneity]\label{eq:homogeneity}

  \begin{equation}
          \Dec(\reda \cdot \Z_x) \approx \reda \cdot \Dec(\Z_x), \quad \forall \Z_x \in \mathcal{Z}, \reda \in \R
  \end{equation}
  \end{property}

\begin{property}[Additivity]\label{eq:additivity}
  \begin{equation}
          \Dec(\Z_u + \Z_v) \approx \Dec(\Z_u) + \Dec(\Z_v), \quad \forall \Z_u, \Z_v \in \mathcal{Z}
  \end{equation}
  \end{property}

  We postulate that approximate linearity can be achieved through a simple data augmentation scheme, without the need to modify the model or loss function. 
  The intuition is that by performing latent space operations (scaling and addition) and asking the decoder to recover the corresponding audio signal, it learns to behave linearly.  Our approach is, in theory, model-agnostic and can be applied to any \ac{AE}. In this work, we apply it to a \ac{CAE} architecture for audio \cite{pasiniMusic2LatentConsistencyAutoencoders2024}, which offers $64 \times$ compression and high-quality reconstruction with a single training objective.

\subsection{Model architecture}\label{sec:model_architecture}

Our model (\texttt{Lin-CAE}) follows the Music2Latent architecture \cite{pasiniMusic2LatentConsistencyAutoencoders2024}, illustrated in Figure  \ref{fig:diagram}(\subref{fig:part1}). We recall the main components here for clarity, but we refer the reader to \cite{pasiniMusic2LatentConsistencyAutoencoders2024} for the full details.

\vspace{\sectionsep}

\noindent \textbf{\ac{CAE} Denoiser representation space}: We use a complex-valued \ac{STFT} followed by the invertible amplitude scaling $\amplitudetransform: \C \to \C  = \beta|c|^\alpha e^{i\angle(c)}$ (with $\alpha =0.65 $ and $\beta = 0.34$) from previous work in complex STFT diffusion \cite{richterSpeechEnhancementDereverberation2023,zhuEdmsoundSpectrogramBased2023,pasiniMusic2LatentConsistencyAutoencoders2024}, which scales the amplitudes in the STFT to roughly $[-1, 1]$ while boosting the high frequencies. We define $\T(x) = \amplitudetransform(\operatorname{STFT}(x)) $ as the transform which maps from waveform to \ac{CM} input space.

\vspace{\sectionsep}

\noindent \textbf{Decoder:} $\Dec$ is composed of a U-Net-based denoiser $\denoiser$ and the inverse transform $\T^{-1}(\X)$. The denoiser reconstructs $\X = \T(x)$ from a noise-corrupted version $\X_\sigma = \X + \sigma \cdot \EPS$ ($\EPS \sim \mathcal{N}(0, I)$), conditioned on the latent $\Z_x$ and the noise level $\sigma$. 
 $\Z_x$ is upsampled with a dedicated network mirroring the U-Net's upsampling block and added to each layer of the U-Net. Noise level information is added to every layer via positional embeddings. $\denoiser$ is parameterized by a noise prediction network $\network$ with a skip connection weighted by coefficients ($c_{\text{skip}},c_{\text{out}}$) that depend on $\sigma$ \cite{karrasElucidatingDesignSpace2022}, which additionally enforces the boundary condition necessary for \ac{CT} \cite{songConsistencyModels2023, songImprovedTechniquesTraining2023, pasiniMusic2LatentConsistencyAutoencoders2024}:


\begin{equation}\label{eq:consistency_denoiser}
  \denoiser(\X_\sigma, \sigma, \Z_x) = c_{\text{skip}}(\sigma)\X_\sigma + c_{\text{out}}(\sigma)\network(\X_\sigma, \sigma, \Z_x)
\end{equation}


During inference, decoding starts from pure noise and the clean signal is reconstructed in a single step conditioned on $\Z_x$.


\vspace{\sectionsep}

\noindent \textbf{Encoder:} $\Enc$ consists of the amplitude transform $\T$ followed by a network that mirrors the U-Net's downsampling blocks.

\subsection{Learning Linearity in CAEs via Implicit Regularization} \label{sec:implicit_linearity}

\noindent \textbf{Implicit homogeneity:} Inspired by \cite{kouzelisEQVAEEquivarianceRegularized2025}, we apply a random positive gain ($\reda$) to the latent and task the decoder to denoise a scaled version of the input ($\reda \cdot x$).  The adapted \ac{CT} training, depicted in Figure \ref{fig:diagram}(\subref{fig:part2}), uses two parallel pathways. $\Enc$ encodes unscaled $x$ to obtain $\Z_x$. Then, the denoiser ($\denoiser$) receives a noisy scaled input ($\T (\reda \cdot x) + \sigma \cdot \EPS$) conditioned on the scaled latent ($\reda \cdot \Z_x$). $\reda$ is never provided as an explicit input to any part of the model, so $\Dec$ must learn to infer the correct output scale from the magnitude of the conditioning latent. If $\reda = 1$, we recover the original \ac{CAE} training.

\vspace{\sectionsep}

\noindent \textbf{Implicit additivity:}   Inspired by Mixit \cite{wisdomUnsupervisedSoundSeparation2020}, we augment our data by creating artificial mixes from pairs of elements $(u, v)$ randomly selected from the training set. When decoding, we replace the true latent of the artificial mix ($\Enc(u+v)$) with the average of the latents of the individual signals: $ \blue{\Z'} \gets   0.5 \cdot (\Z_u + \Z_v)$, where 
$(\Z_u= \Enc(u), \Z_v=\Enc(v))$, as illustrated in Figure \ref{fig:diagram}(\subref{fig:part3}). The denoiser is thus conditioned on $\blue{\Z'}$ and tasked with denoising  $\T(\reda \cdot 0.5 \cdot (u+v)) +\sigma \cdot \EPS)$. We augment the training batch with mixtures by concatenating it with a version of itself that has been circularly shifted by one position and averaged with the original (Figure \ref{fig:diagram}(\subref{fig:part4})).






\vspace{-0.2cm}

\subsection{Training}

 With the modifications described in Section \ref{sec:implicit_linearity} highlighted in color, we train the model under the \ac{CT} objective $\mathcal{L}_{\text{CT}}$ \cite{songConsistencyModels2023,songImprovedTechniquesTraining2023,pasiniMusic2LatentConsistencyAutoencoders2024}, which minimizes the distance between the denoised output of a teacher ($f_{\theta^-}$) and student ($f_{\theta}$) models ($\theta^- \gets \text{stopgrad}(\theta)$) for noisy inputs at two different noise levels, $\sigma_{t_1}$ and $\sigma_{t_2}$.



\begin{equation}
  \begin{split}
    \mathcal{L}_{\text{CT}}  = \mathbb{E}_{\X, \EPS, t_{2}, \reda} \bigg[ \lambda \cdot d\Big(  f_{\theta}&(\T(\reda \cdot x) + \sigma_{t_2} \cdot  \EPS, \sigma_{t_2}, \reda \cdot \blueZ'), \\
     f_{\theta^-}&(\T(\reda \cdot x) + \sigma_{t_1}\cdot  \EPS, \sigma_{t_1}, \reda \cdot \blueZ') \Big) \bigg],
  \end{split}
  \end{equation}
  

\noindent where $\blueZ'$ is either $\Z_x$ or $0.5 \cdot (\Z_u + \Z_v)$ depending on whether the input waveform is sampled from the training set or is an artificial mixture ($x=0.5 \cdot (u + v$)), $\EPS \sim \mathcal{N}(0, I)$ is a shared noise direction, 
 $\lambda(\sigma_{t_1}, \sigma_{t_2})  = \frac{1}{\sigma_{t_2} - \sigma_{t_1}}$ is a weighting function designed to give higher weights to smaller noise increments \cite{songImprovedTechniquesTraining2023}. We sample time steps $t_2 \in [0, 1]$ and corresponding noise levels $\sigma_{t_1, t_2}$ following \cite{pasiniMusic2LatentConsistencyAutoencoders2024}, with $t_1 = \max(t_2 - \Delta_{t_k}, 0)$, where $\Delta_{t_k}$ depends on the training step $k$ and decays over training ($\Delta t_k = \Delta t_0^{\frac{k}{K}(e_K - 1) + 1}$, $\Delta t_0=0.1$, $e_K=2.0$). $d$ is the Pseudo-Huber loss \cite{charbonnierDeterministicEdgepreservingRegularization1997}: $d(x, y) = \sqrt{|x - y|^2 + c^2} - c$, with $c=5.4e-4$.

\vspace{\sectionsep}

  \noindent \textbf{Training details: } We train all models for $800$K steps with a batch size of $20$ before mixture creation (final batch size of $40$). Training takes $\approx$ $8$ days on $1$ L$40$S GPU. The models are optimized with RAdam with learning rate $\in [10^{-4}, 10^{-6}]$ (linear warmup for $10$K steps, cosine decay for the rest). We track an \ac{EMA} of the model parameters $\theta$ to be used for inference, updated every $10$ steps.

  Homogeneity gains are sampled from a uniform distribution in $[a_\text{min}, a_\text{max}]$ and applied with probability $0.8$ to each sample in the batch. If $|a|<0.05$, we set it to $0$. Gains are clipped so that the maximum absolute value of the waveform does not exceed $1$. We anneal gain range from $(a_\text{min}=0, a_\text{max}=3)$ to no gain $(1, 1)$ over training using a piecewise cosine schedule. 
   Gain boundaries are defined as: $a_\text{min}(k) = 1 - \cdot C(k)$ and $a_\text{max}(k) = 1 + 2 \cdot C(k)$, where $C(k)$ is $1$ if $k \le 0.2K$, $0$  if $k \ge 0.9K$, and follows a cosine decay $C(k) = \frac{1}{2}(1 + \cos(\pi \cdot \frac{k - 0.2K}{0.7K}))$ for the intermediate steps.

\section{Experiments and Results}

Results are displayed in Tables \ref{tab:add_sep} and  \ref{tab:musiccaps_metrics}. Best results are in bold, second best are underlined, and arrows indicate whether higher ($\uparrow$) or lower ($\downarrow$) is better.
 Audio examples and supplementary material are available online \footnote{{\scriptsize \url{https://bernardo-torres.github.io/projects/linear-cae}}}.
 

\begin{figure}
  \centering
  \includegraphics[width=0.7\columnwidth, trim={5 10 5 0}]{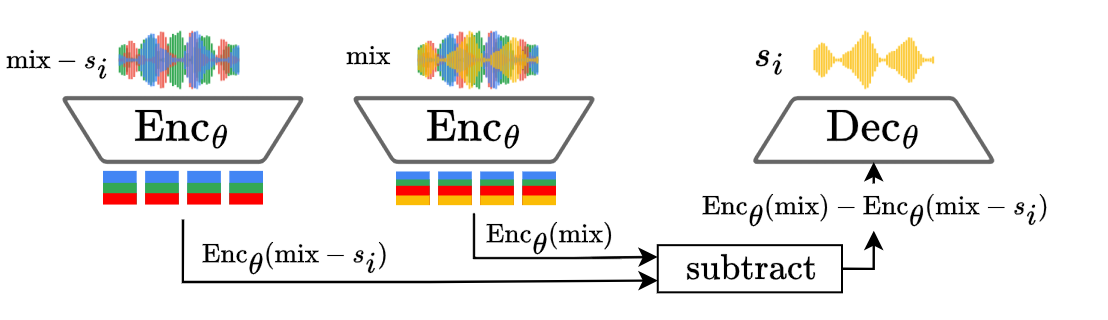}
  \caption{Oracle Music Source Separation via latent arithmetic.}
  \label{fig:ss}
\end{figure}

\subsection{Experimental setup}

\noindent\textbf{Data:} The training corpus is a large-scale music/speech dataset compiled from MTG-Jamendo \cite{bogdanovMTGjamendoDatasetAutomatic2019}, MoisesDB \cite{pereiraMoisesDBDatasetSource2023}, M4Singer\cite{zhangM4SingerMultistyleMultisinger2022}, DNS-Challenge \cite{dubeyICASSP2023Deep2023}, and E-GMD \cite{callenderImprovingPerceptualQuality2020}. Tracks are sampled with weights $(60, 20, 9, 8, 3)$ respectively, and a segment of $2$ seconds is randomly cropped from each track. Segments are converted to mono by averaging stereo channels and resampled to $44.1$ kHz.

\vspace{\sectionsep}

\noindent\textbf{Baselines:} We report metrics using the publicly available weights of Music2Latent (\texttt{M2L-Pub}) \cite{pasiniMusic2LatentConsistencyAutoencoders2024} . As a fairer baseline, we retrain \texttt{M2L-Pub}~on the same data as our model, including the random gains and artificial mixtures for data augmentation, but without our proposed implicit homogeneity and additivity conditioning strategies (\texttt{M2L}). We also report metrics for the public weights from Stable Audio $1.0$ VAE \cite{evansStableAudioOpen2025} (\texttt{SA-VAE}), serving as another autoencoder baseline with a different architecture and training procedure.

\input{tables/add_sep.tex}
\input{tables/musiccaps_metrics.tex}

\subsection{Reconstruction Quality}

We evaluate the reconstruction quality of the \acp{AE} on 2-second audio chunks ($0.5$s overlap) from the MusicCaps dataset \cite{agostinelliMusicLMGeneratingMusic2023}. We reconstruct the full track using overlap-add and compute three metrics: \begin{enumerate*}[label=(\Roman*)]
\item The \ac{SNR} on the waveform. 
\item A \ac{MSS} between the original and reconstructed log-mel spectrograms\footnote
{We use mel-spectrogram distances computed over multiple resolutions ($80$ mel bands and hop lengths of $\approx$ $10$, $25$, and $50$ ms) as a proxy for a phase-insensitive perceptual metric.  }. 
\item The \ac{KAD} \cite{chungKADNoMore2025} between the embedding distributions of the original and reconstructed tracks\footnote{ Embedding distribution metrics have become a standard way of evaluating generative models, so we include it for completeness. KAD is computed using half of the reconstructed tracks, using the other half as the reference set. We use the \texttt{kadtk} library with the LAION-CLAP model.}. 
\end{enumerate*}

Table \ref{tab:musiccaps_metrics} summarizes the results. Our retrained baseline (\texttt{M2L}) already improves on the public \texttt{M2L-Pub} weights \cite{pasiniMusic2LatentConsistencyAutoencoders2024}, likely due to our distinct training data, batch size and augmentations. The linearized model (\texttt{Lin-CAE}) remains notably comparable on \ac{MSS} and improves further on \ac{SNR}, indicating that our method does not degrade reconstruction fidelity. Compared to \texttt{SA-VAE},  \texttt{Lin-CAE} is slightly worse, a similar finding to \cite{pasiniMusic2LatentConsistencyAutoencoders2024,pasiniMusic2Latent2AudioCompression2025}. We note that \texttt{SA-VAE} uses different data and a two-stage  training procedure that optimizes both  reconstruction and adversarial losses, while  \texttt{Lin-CAE} is trained with a single denoising objective. KAD scores for  \texttt{Lin-CAE} are better than our retrained \texttt{M2L-Pub} model and \texttt{SA-VAE} , but are worse than the public \texttt{M2L} weights (possibly reflecting the different training data distribution). 

\vspace{ -\sectionsep}

\subsection{Homogeneity}

We use the same data and setup as for reconstruction, but we additionally assess the model's ability to preserve scaling by drawing a random scalar $\reda$ for each track $x$ and measuring the effect on the output of both the encoder and decoder. We measure \begin{enumerate*}[label=(\Roman*)]
\item \textbf{Encoder Homogeneity (Enc-Hom.)} as the relative $L_2$ error between the scaled input's latent and the latent of the scaled input: $\| \Enc(\reda \cdot x) - \reda \cdot \Enc(x) \|_2 / \| \reda \cdot \Enc(x) \|_2$. We employ a relative error to normalize for the smaller latent norms produced by models trained with homogeneity, which would make a standard $L_2$ comparison misleading;
\item \textbf{Decoder Homogeneity (Dec-Hom.)} as the SNR and MSS between  $\reda \cdot \Dec(\Z_x)$ and $\Dec(\reda \cdot \Z_x)$. 
\end{enumerate*}

Results are shown in Table \ref{tab:musiccaps_metrics}, where it is shown that
\texttt{Lin-CAE} achieves much better \textbf{decoder homogeneity }(Table \ref{tab:musiccaps_metrics})  properties compared to both baselines. For all baselines, we see a significant degradation in reconstruction metrics when testing the effect of random gains, while our model is much more robust. A high degradation in the log-domain \ac{MSS} indicates that the scaling of the latent translates to significant timbral changes in the output. \texttt{Lin-CAE}'s MSS degrades less ($1.01 \to 1.37$) than the baselines (\texttt{M2L}: $0.98 \to 2.27$ and \texttt{SA-VAE}: $0.72 \to 3.03$). The \textbf{encoder} also exhibits significantly lower homogeneity errors compared to all baselines.

\vspace{-\sectionsep}

\subsection{Additivity}

Additivity is evaluated on four source mixtures from the Musdb18-HQ test set \cite{rafiiMUSDB18HQUncompressedVersion2019}, where a mixture is : $\text{mix} = \sum^4_{i=1} s_i$. $s_i$ correspond to either vocals, bass, drum or other. 
We measure \textbf{Encoder Additivity Error} as the deviation of a mixture's latent from the sum of its source latents, calculated as the relative $L_2$ error: $\| \Enc(\text{mix}) - \sum_i \Enc(s_i) \|_2 / \| \Enc(\text{mix}) \|_2$. For \textbf{Decoder Additivity}, we test the ability to reconstruct a mixture from the sum of its source latents by \ac{SNR}/\ac{MSS}  between the decoded sum of latents, $\Dec(\sum_i \Enc(s_i))$, and the \textit{autoencoded} mixture, $\Dec(\Enc(\text{mix}))$.

 Our results (Table \ref{tab:add_sep}) show that \texttt{Lin-CAE}  achieves a very low MSS ($0.99$) compared to the baselines ($>5$), indicating that the summation in the latent space is very close to the autoencoded mixture, while the baselines produce a significantly degraded output. We encourage the reader to listen to the audio examples, as this is very clearly perceptible. The \textbf{encoder} also exhibits significantly lower additivity errors. Paired with the results from encoder homogeneity, this suggests that our implicit conditioning strategy encourages the entire \ac{AE} mapping to become more linear, not just the decoder.

\vspace{ -\sectionsep}
\vspace{ -\sectionsep}

\subsection{Music source separation via latent arithmetic}

We perform source separation via latent arithmetic on the Musdb18-HQ test set. 
 A source $\hat{s}_i$ is estimated by subtracting the latent representation of its accompaniment from the latent of the full mixture: $\hat{s}_i = \Dec(\Enc(\text{mix}) - \Enc(\text{mix} - s_i))$. Figure \ref{fig:ss} illustrates this process. The quality of the separated source is evaluated against the \textit{autoencoded} ground-truth source, $\Dec(\Enc(s_i))$, using \ac{SI-SDR} \cite{lerouxSDRHalfbakedWell2019} and \ac{MSS}.

Table \ref{tab:add_sep} shows that  \texttt{Lin-CAE} significantly outperforms the baselines in latent arithmetic/oracle source separation tasks, achieving higher \ac{SI-SDR} and lower \ac{MSS} scores by a large margin for every instrument. The model was not trained to recover the exact phase, which is why the relatively low SDR values are observed for both reconstruction and separation. Notably, however, both separation metrics are of the same order of magnitude as the reconstruction values for the mix, which can be seen as an upper bound for the model. 
 That is not the case, however, for any of the baselines, with the least significant gap for \texttt{SA-VAE}. We note that during training, we have only shown positive gains to the model, while during separation, we are effectively applying negative gains when subtracting latents. This further indicates the robustness of the learned linear structure.

\textbf{Ablations:} We retrain  \texttt{Lin-CAE} with only one of the two properties (homogeneity or additivity) and report source separation results in Table \ref{tab:add_sep}. Performance in every metric degrades significantly when only one of the two properties is applied. Surprisingly, training with only homogeneity seems to help the additivity and source separation much more as a byproduct than training with only additivity. 

\vspace{-0.3cm}

\section{Conclusion}

We introduce a simple method to induce an approximately linear latent space in audio autoencoders during training. Using implicit conditioning via data augmentation, our approach enforces homogeneity and additivity (properties of a linear map) without modifying the architecture or objective. Experiments with Consistency Autoencoders show that linearization preserves high-quality reconstruction and can be used for source separation via latent arithmetic.

Our approach enables more interpretable and controllable audio generation. Future work could extend this method to improved \ac{CAE} architectures \cite{pasiniMusic2Latent2AudioCompression2025} and downstream tasks such as generative source separation \cite{marianiMultisourceDiffusionModels2024}. Direct, low-level manipulations in compressed space promise new possibilities for audio editing and efficient signal processing.

\apptocmd{\thebibliography}{\footnotesize}{}{}
\bibliographystyle{IEEEbib}
{
\bibliography{macros, bernardo}}

\end{document}

%% file: tables/add_sep.tex
\begin{table*}
\caption{Additivity and oracle Music Source Separation using latent space arithmetic  on the MUSDB18-HQ dataset.}
\vspace{-6pt}
\label{tab:add_sep}
\centering
\small
\setlength{\tabcolsep}{2pt} 

\begin{tabular}{llllllllllllll}
\toprule
 & \multicolumn{3}{c}{Additivity} & \multicolumn{2}{c}{Reconstruction} & \multicolumn{8}{c}{Separation} \\
\cmidrule(lr){2-4} \cmidrule(lr){5-6} \cmidrule(lr){7-14}
 & \multicolumn{2}{c}{Decoder} & Encoder & \multicolumn{2}{c}{Mix} & \multicolumn{2}{c}{Bass} & \multicolumn{2}{c}{Drums} & \multicolumn{2}{c}{Other} & \multicolumn{2}{c}{Vocals} \\
\cmidrule(lr){2-3} \cmidrule(lr){4-4} \cmidrule(lr){5-6} \cmidrule(lr){7-8} \cmidrule(lr){9-10} \cmidrule(lr){11-12} \cmidrule(lr){13-14}
Model & MSS $\downarrow$ & SNR $\uparrow$ & Error $\downarrow$ & MSS $\downarrow$ & SNR $\uparrow$ & MSS $\downarrow$ & SI-SDR $\uparrow$ & MSS $\downarrow$ & SI-SDR $\uparrow$ & MSS $\downarrow$ & SI-SDR $\uparrow$ & MSS $\downarrow$ & SI-SDR $\uparrow$ \\
\midrule
\texttt{M2L-Pub} \cite{pasiniMusic2LatentConsistencyAutoencoders2024} & $5.01$ & $-0.79$ & $2.82$ & $1.16$ & $-1.29$ & $95.34$ & $-14.85$ & $28.08$ & $-16.30$ & $6.78$ & $-16.11$ & $5.10$ & $-18.16$ \\
\texttt{SA-VAE} \cite{evansStableAudioOpen2025} & $5.38$ & $-12.58$ & $1.71$ & $\mathbf{0.57}$ & $\mathbf{4.03}$ & $323.26$ & $\underline{-2.81}$ & $87.80$ & $\underline{-3.11}$ & $3.65$ & $\underline{-4.10}$ & $3.08$ & $\underline{-4.48}$ \\
\texttt{M2L} & $5.21$ & $-0.48$ & $2.73$ & $0.97$ & $0.10$ & $39.56$ & $-11.11$ & $12.26$ & $-11.95$ & $5.30$ & $-11.81$ & $3.49$ & $-12.56$ \\
\rowcolor{gray!20} \texttt{Lin-CAE} & $\mathbf{0.99}$ & $\mathbf{1.22}$ & $\mathbf{0.60}$ & $1.00$ & $0.21$ & $\mathbf{1.58}$ & $\mathbf{-0.59}$ & $\mathbf{1.16}$ & $\mathbf{-0.28}$ & $\mathbf{1.21}$ & $\mathbf{-1.82}$ & $\mathbf{1.23}$ & $\mathbf{-1.18}$ \\
\cmidrule(lr){1-14}
\multicolumn{14}{l}{\scriptsize\textit{Ablations}} \\
 - Additivity & $\underline{1.68}$ & $\underline{0.30}$ & $2.07$ & $1.39$ & $-0.06$ & $\underline{6.61}$ & $-7.12$ & $\underline{2.37}$ & $-6.11$ & $3.44$ & $-8.52$ & $\underline{2.87}$ & $-12.18$ \\
 - Homogeneity & $4.24$ & $-0.89$ & $\underline{1.42}$ & $\underline{0.96}$ & $\underline{0.43}$ & $23.61$ & $-5.96$ & $7.84$ & $-5.14$ & $\underline{2.98}$ & $-8.30$ & $3.08$ & $-7.95$ \\
\bottomrule
\vspace{-6pt}
\end{tabular}
\end{table*}

%% file: tables/musiccaps_metrics.tex
\begin{table}
\caption{MusicCaps Reconstruction and Homogeneity (Hom.).}
\vspace{-6pt}
\label{tab:musiccaps_metrics}
\centering
\small
\setlength{\tabcolsep}{1.5pt} 

\begin{tabular}{lcccccc}
\toprule
 & \multicolumn{2}{c}{Reconstruction} & \multicolumn{2}{c}{Dec-Hom.} & Enc-Hom. &   \\
\cmidrule(lr){2-3} \cmidrule(lr){4-5} \cmidrule(lr){6-6}
Model & MSS $\downarrow$ & SNR $\uparrow$ & MSS $\downarrow$ & SNR $\uparrow$ & Error $\downarrow$ & KAD $\downarrow$ \\
\midrule
\texttt{M2L-Pub} \cite{pasiniMusic2LatentConsistencyAutoencoders2024} & $1.14$ & $1.85$ & $2.52$ & $-4.69$ & $12.13$ & $\mathbf{5.69}$ \\
\texttt{SA-VAE} \cite{evansStableAudioOpen2025} & $\mathbf{0.72}$ & $\mathbf{7.32}$ & $3.03$ & $\underline{-1.27}$ & $\underline{4.59}$ & $6.27$ \\
\texttt{M2L} & $\underline{0.98}$ & $3.09$ & $\underline{2.27}$ & $-2.30$ & $8.52$ & $6.53$ \\
\rowcolor{gray!20} \texttt{Lin-CAE} & $1.01$ & $\underline{3.19}$ & $\mathbf{1.37}$ & $\mathbf{0.86}$ & $\mathbf{0.69}$ & $\underline{6.19}$ \\
\bottomrule
\vspace{-6pt}
\end{tabular}
\end{table}